\documentclass{ieeeaccess}

\usepackage{cite}
\usepackage{amsmath,amssymb,amsfonts}
\usepackage{bbold}
\usepackage{algorithmic}
\usepackage{graphicx}
\usepackage{textcomp}
\usepackage{trimclip}
\usepackage{soul}

\usepackage{mathtools, cuted}
\usepackage{float}
\usepackage{array}
\usepackage{multirow}
\usepackage{braket}
\usepackage{url}
\usepackage{algorithmic}
\usepackage{booktabs}
\usepackage{colortbl}
\usepackage{xspace}
\usepackage{resizegather} 
\usepackage{mathrsfs}

\usepackage[breaklinks,colorlinks=true,linkcolor=black,anchorcolor=black,citecolor=black,filecolor=black,menucolor=black,runcolor=black,urlcolor=black]{hyperref}
\usepackage{hyphenat}
\usepackage{microtype}
\emergencystretch=\maxdimen
\usepackage[caption=false]{subfig}
\usepackage{cancel}

\usepackage[acronym]{glossaries}
\glsdisablehyper
\makeglossaries
\newacronym{pca}{PCA}{principal component analysis}
\newacronym{qnn}{QNN}{quantum neural network}
\newacronym{qml}{QML}{quantum machine learning}
\newacronym{awgn}{AWGN}{additive white Gaussian noise}
\newacronym{snr}{SNR}{Signal-to-Noise Ratio}
\newacronym{ser}{SER}{Symbol Error Rate}
\newacronym{ber}{BER}{Bit Error Rate}
\newacronym{mimo}{MIMO}{Multiple Input Multiple Output}
\newacronym{psk}{PSK}{Phase Shift Keying}
\newacronym{bpsk}{BPSK}{Binary Phase Shift Keying}
\newacronym{qpsk}{QPSK}{Quadrature Phase Shift Keying}
\newacronym{qam}{QAM}{Quadrature Amplitude Modulation}
\newacronym{4qam}{4-QAM}{4-level Quadrature Amplitude Modulation}
\newacronym{16qam}{16-QAM}{16-level Quadrature Amplitude Modulation}
\newacronym{qaoa}{QAOA}{Quantum Approximate Optimization Algorithm}

\definecolor{Orange}{rgb}{1,0.65,0}
\definecolor{Red}{rgb}{1,0,0}
\definecolor{Blue}{rgb}{0, 0, 1}
\definecolor{Emerald}{rgb}{0.31, 0.78, 0.47}
\definecolor{NavyBlue}{rgb}{0, 0, 0.5}
\definecolor{LimeGreen}{rgb}{0.494, 0.827, 0.129}
\definecolor{bondiblue}{rgb}{0.0, 0.58, 0.71}
\definecolor{darkgray}{rgb}{0.66, 0.66, 0.66}
\definecolor{lightgray}{rgb}{0.33, 0.33, 0.33}
\definecolor{bazaar}{rgb}{0.6, 0.47, 0.48}
\definecolor{blue-violet}{rgb}{0.54, 0.17, 0.89}
\definecolor{brightmaroon}{rgb}{0.76, 0.13, 0.28}
\definecolor{rev}{rgb}{0.96, 0.32, 0.12}
\definecolor{revv}{rgb}{0.83, 0.24, 0.09}

\def\BibTeX{{\rm B\kern-.05em{\sc i\kern-.025em b}\kern-.08em
    T\kern-.1667em\lower.7ex\hbox{E}\kern-.125emX}}

\begin{document}
\bstctlcite{IEEEexample:BSTcontrol} 

\newcommand{\etal}{\textit{et al.}}
\newcommand{\symspace}{[1, ...,\! \, M]}
\newcommand{\iqspace}{\mathbb{R}^{2n}}
\newcommand{\tx}{\mathcal{T\!X}}
\newcommand{\ch}{\mathcal{C\!H}}
\newcommand{\rx}{\mathcal{R\!X}}
\newcommand{\sym}{s}
\newcommand{\estsym}{\hat{s}}
\newcommand{\xsingal}{\mathbf{x}}
\newcommand{\ysingal}{\mathbf{y}}
\newcommand{\noise}{\mathbf{n}}
\newcommand{\chvec}{\mathbf{h}}

\newcommand{\qsyle}[1]{\mathsf{#1}}
\newcommand{\qnsyle}[1]{\mathsf{#1}}

\newcommand{\hybrid}{\qsyle{H}\text{-}\qsyle{C\!A\!E}}

\newcommand{\ccae}{\qsyle{C\!C}\text{-}\qsyle{A\!E}}
\newcommand{\cqae}{\qsyle{C\!Q}\text{-}\qsyle{A\!E}}
\newcommand{\qcae}{\qsyle{Q\!C}\text{-}\qsyle{A\!E}}
\newcommand{\qqae}{\qsyle{Q\!Q}\text{-}\qsyle{A\!E}}
\newcommand{\cqrae}{\qsyle{C\!Q}^{\qsyle{R}}\text{-}\qsyle{A\!E}}

\newcommand{\CC}[1]{\qsyle{C\!C}_{\qnsyle{#1}}}
\newcommand{\CQ}[1]{\qsyle{C\!Q}_{\qnsyle{#1}}}
\newcommand{\QC}[1]{\qsyle{Q\!C}_{\qnsyle{#1}}}
\newcommand{\QQ}[1]{\qsyle{Q\!Q}_{\qnsyle{#1}}}
\newcommand{\CQR}[1]{\qsyle{C\!Q}_{\qnsyle{#1}}^{\qnsyle{R}}}
\newcommand{\CQST}[1]{\qsyle{C\!Q}_{\qnsyle{#1}}^{\qnsyle{16}}}
\hyphenation{op-tical net-works semi-conduc-tor}

\newcommand{\papertitle}{Quantum-Classical Autoencoder Architectures for End-to-End Radio Communication}


\history{Date of publication xxxx 00, 0000, date of current version xxxx 00, 0000.}
\doi{10.1109/ACCESS.2021.DOI}
\title{\papertitle}

\author{
\uppercase{Zsolt I. Tabi}\authorrefmark{1,2}, 
\uppercase{Bence Bak\'o}\authorrefmark{1,2},
\uppercase{D\'aniel T. R. Nagy}\authorrefmark{1,2},
\uppercase{P\'eter Vaderna}\authorrefmark{3},
\uppercase{Zs\'ofia Kallus}\authorrefmark{3}, 
\uppercase{P\'eter H\'aga}\authorrefmark{3}, and
\uppercase{Zolt\'an Zimbor\'as}\authorrefmark{1,2}}

\address[1]{E\"otv\"os Lor\'and University, Budapest, Hungary (e-mail: zsolttabi@ik.elte.hu)
}
\address[2]{Quantum Computing and Information Group, HUN-REN Wigner Research Centre for Physics, Budapest, Hungary (e-mail: \{bako.bence, nagy.dani, zimboras.zoltan\}@wigner.hu)
}
\address[3]{Ericsson Research, Budapest, Hungary (e-mail: \{peter.vaderna, zsofia.kallus, peter.haga\}@ericsson.com)
}

\markboth
{Z. I. Tabi \headeretal: \papertitle}
{Z. I. Tabi \headeretal: \papertitle}

\corresp{Corresponding authors: Zsolt Tabi (e-mail: zsolttabi@ik.elte.hu), P\'eter H\'aga (e-mail: peter.haga@ericsson.com), Zolt\'an Zimbor\'as (e-mail: zimboras.zoltan@wigner.hu).}

\begin{abstract}

This paper presents a comprehensive study on the possible hybrid quantum-classical autoencoder architectures for end-to-end radio communication against noisy channel conditions using standard encoded radio signals.
The hybrid scenarios include single-sided, i.e., quantum encoder (transmitter) or quantum decoder (receiver), as well as fully quantum channel autoencoder (transmitter-receiver) systems. 
We provide detailed formulas for each scenario and validate our model through an extensive set of simulations. Our results demonstrate model robustness and adaptability. Supporting experiments are conducted utilizing 4-QAM and 16-QAM schemes and we expect that the model is adaptable to more general encoding schemes. 
We explore model performance against both additive white Gaussian noise and Rayleigh fading models.
Our findings highlight the importance of designing efficient quantum neural network architectures for meeting application performance constraints -- including data re-uploading methods, encoding schemes, and core layer structures.
By offering a general framework, this work paves the way for further exploration and development of quantum machine learning applications in radio communication.

\end{abstract}


\titlepgskip=-15pt
\maketitle

\section{Introduction}\label{sec:introduction}

With artificial intelligence (AI) and machine learning (ML) advancing industrial applications in many areas, there is an active research of emerging technologies to enhance various aspects of their application-level performance. Quantum computing is one of the promising new paradigms, where the high expressivity of quantum neural networks has the potential for efficient representation of complex classical problems. 

\Gls{qml} has emerged as a promising intersection of quantum computing and machine learning~\cite{Schuld2014, PhysRevResearch.1.033063}. \Glspl{qnn}, characterized by unique variational quantum circuits, function as quantum analogues of classical neural networks and can be optimized using gradient-based or gradient-free optimization algorithms, forming hybrid quantum-classical training loops~\cite{Wierichs2022generalparameter, chen2022variational}. 

The landscape of \gls{qnn}-based models include quantum convolutional neural networks~\cite{QCNNWei2022}, generative models~\cite{LLoydQGAN}, long short-term memories~\cite{2021APSQLSTM}, and autoencoders~\cite{PhysRevA.102.032412,Romero_2017,PhysRevLett.124.130502}. 

Variational quantum circuits
are suitable for near-term, noisy quantum compute resources, as opposed to the more distant goal of fault-tolerant quantum computing.
However, these models are still constrained by design factors, e.g., the limited number of qubits available. 
Consequently, these parameterized circuits are often results of a delicate balance: minimizing circuit size and gate count, while maximizing the expressivity of efficiently trainable models. 
This leads to an empirically driven approach, where experimental results guide the architecture design process~\cite{schuld2018supervised}.

The use of a hybrid classical-quantum autoencoder was explored by Sakhnenko~A.~\etal \cite{sakhnenko2022hybrid} for anomaly detection tasks.
They demonstrated that incorporating a Parameterized Quantum Circuit (PQC) into the bottleneck layer of a classical autoencoder significantly enhances its performance.
The hybrid model showed improvements in key metrics such as precision, recall, and F1 score.
This study highlights the potential of quantum circuits to augment classical machine learning models, particularly in applications that require high-dimensional data compression and feature extraction.

Wireless communications systems face many challenges as they aim at reliable communication while maintaining high throughput. Traditional detection algorithms have been successfully used for fast decoding of the transmitted signals, often modeling noisy channels by statistical characterization of various environments~\cite{7c51be827a654249a0fdbd26fcd4c47a,meyr1998digital}. In the R\&D efforts towards the sixth generation (6G) of mobile cellular networks, extreme performance of radio access technology is targeted to serve the diverse requirements of Industry and Society~\cite{ericsson6G}.
The design of mobile network infrastructure is including AI and ML solutions to improve performance, flexibility, and efficiency~\cite{HexaX}. ML algorithms are introduced into different components, i.e., network management and control, analytics, and even in the physical layer.  In novel radio communication concepts, the receive or transmit side, or even end-to-end learning techniques can be applied for joint optimization to counteract radio signal degradation.
Representing the end-to-end communication system as a deep Channel Autoencoder was introduced in Refs.~\cite{7886039,8054694}, and the model was further studied for various applications in Refs.~\cite{dorner2017deep,o2017deep,o2017physical,zou2021channel,GARCIA2022192,9446711,9204706,9154335}.

In a recent work~\cite{tzs20222hybrid}, a hybrid quantum-classical extension was presented for this Channel Autoencoder. A single-sided hybrid model, using a \gls{qnn} only on the receiver side was introduced. This approach with end-to-end training for decoding achieved performance respecting the time constraints of standard radio communication. In Ref.~\cite{rathi2023quantum}, quantum channels replaced classical radio communication to demonstrate a Channel Autoencoder for quantum communication. 
This approach is fundamentally different from binary optimization problems, where one can also leverage quantum computing resources in telecommunication domain \cite{marosits2021exploring, tzs2021evaluation, kim2019leveraging}.


To address complexity in end-to-end radio communication, here we study the viability of leveraging \glspl{qnn} in Channel Autoencoders, since they provide efficient representation of high-dimensional, noisy, and non-linear patterns typical in industrially relevant problems. In this paper, we propose to benchmark various architectures through some standard simulated scenarios. We also consider the further limiting time constraint of radio communication applications, hence optimizing circuit design not only for faster training, but also for shorter \glspl{qnn} inference time.


The organization of this paper is as follows: Sec.~\ref{sec:background} offers an overview of \glspl{qnn} and discusses the Channel Autoencoder model abstraction of an end-to-end radio communication line. In Sec.~\ref{sec:qpae}, we present our hybrid Channel Autoencoder model architectures, which allow for \glspl{qnn} implementations in transmitter and receiver sides. Subsequently, in Sec.~\ref{sec:experiments}, we evaluate the performance of our hybrid models on various problems. Finally, in Sec.~\ref{sec:summary}, we summarize our findings.


\section{Technical Background}\label{sec:background}
In this section, we review the relevant theoretical concepts. First, we introduce the the basic building blocks of QNN models, concentrating on supervised learning tasks with classical data. Then we continue with the introduction of classical autoencoders and definition of our problem of interest. Finally, we show how this problem can be posed as a learning task and solved with classical autoencoders.

\subsection{Building Blocks of Quantum Neural Networks}

In the context of ML with classical data, a \gls{qnn} has three main components: qubit encoding for embedding the classical data, the parameterized layers, and the measurement.
The loss function in such a model is usually expressed as:
\begin{equation}
    f_{\theta}(x) = \bra 0 U^{\dagger}(x,\theta) O U(x,\theta) \ket 0 \enspace,
\end{equation}
\noindent where $U(x,\theta)$ represents the unitary operator of the quantum circuit consisting of both encoding the input data $x$ and applying a set of parametrized gates, with trainable parameters $\theta$, and $O$ is a chosen observable to be measured.

The first building block of a \gls{qnn} is the \textit{data encoding}, i.e., the representation of classical data on a quantum computer.
Choosing an appropriate data encoding is critical, since it is considered to be one of the possible bottlenecks of \glspl{qnn}, especially for near-term devices~\cite{schuld2021machine}.

In this work, we consider multiple approaches, all of which are sub-classes of either \textit{basis} or \textit{time-evolution} encoding.
Basis encoding directly encodes classical data by associating the state of each qubit with a bit from the binary representation of the input feature.
Time-evolution encoding, on the other hand, is an indirect way of encoding feature vectors.
In general, one can encode a scalar $t$ through a unitary transformation $U(t) = \exp{(-itH)}$.
Using only single-qubit Pauli operators in $H$, we arrive at the \textit{angle encoding} method, which is the shallowest subclass of this strategy, however, we also consider more complex Hamiltonians for this task.

Another key aspect of data encoding methods is the way they affect the expressivity of the quantum circuit. In Ref.~\cite{PerezSalinas2020DataRF} the authors introduced the \textit{data re-uploading trick}, which repeats the encoding block before each variational layer.
The intuition behind the effectiveness of this method is that by re-introducing the input before each layer, one can mimic the computational structure of typical classical deep neural networks, where the copying of the classical information is readily available, which would be, without this trick, prohibited by the no-cloning theorem in \gls{qml}.
In Ref.~\cite{PhysRevA.103.032430}, the authors have shown that re-introducing the classical information sequentially (data re-uploading) and in parallel on different subsystems are both beneficial for the expressivity and performance of the model,
hence, in this work, we also make use of these methods.

The second building block of a \gls{qnn} is the \textit{variational layer}, consisting of well chosen parametrized  circuit templates, e.g., weakly and strongly entangling layers \cite{PhysRevA.101.032308}. Thus, besides the encoding operators, the final Ansatz consists of the repeated application of the QNN layers $U_k(\theta_k)$ with independent trainable parameters as
\begin{equation}
    U(\theta) = U_L(\theta_L) \dots U_2(\theta_2)U_1(\theta_1).
\end{equation}

The final building block is the \textit{measurement}, from which the loss function value can be inferred. This can be based on the expectation value of some -- preferably local -- observable, or the outcome probabilities of the computational basis measurements. While both of these methods are often used, it is worth mentioning, that the number of measurement samples needed to estimate outcome probabilities in general grows exponentially with the system size. Contrary to this, only polynomial number is needed for estimating the expectation values of local observables. 

\subsection{Autoencoders}

An autoencoder is an unsupervised learning technique, often implemented using neural networks, that efficiently compresses data by learning to approximate its original input from a compressed state, also known as the latent space~\cite{ng2011sparse, Hinton2006, Goodfellow-et-al-2016, NejiHala8893133}. Autoencoders are typically represented as a pair of functions: an encoder, which learns a latent space representation of its inputs, $\mathbf{h}=f(\mathbf{x})$, and a decoder, which attempts to reconstruct an approximation of the input from the latent space, $\mathbf{\tilde{x}}=g(\mathbf{h})$.

One of the key properties of autoencoders is their ability to construct a \emph{meaningful} representation of input data points, similar to \gls{pca}. To ensure that the autoencoder learns the most important features of the input data, which could be very high-dimensional, it defines a latent space that is undercomplete, i.e., has a dimension less than the input space.

Autoencoders can be trained using a loss function of the form $L(\mathbf{x}, \mathbf{\tilde{x}})$, which measures the dissimilarity between the input and its reconstruction. If $L$ is chosen to be nonlinear, the autoencoder can be considered as a nonlinear generalization of \gls{pca}.

\subsection{Wireless Communication and Channel Models}

Wireless communication systems, in their simplest form, can be modeled by three main components: a transmitter, a channel, and a receiver. The flow of information in the system is as follows: We have a set of $M$ possible symbols $\sym \in \symspace$ to be transmitted. When the transmitter wishes to send a symbol, it applies the transformation $\tx:\symspace\mapsto\iqspace$, where $n$ is the number of discrete \emph{channel uses}. The transmitted signal, $\xsingal=\tx(\sym)$, is a complex vector with $2n$ components, normalized to average power (although some hardware implementations might require other normalization methods). The channel applies a probabilistic transformation to the transmitted signal $\ch:\iqspace\mapsto\iqspace$, which could be additive noise (\gls{awgn}), or some other impairment, such as Rayleigh fading. Once the received signal $\ysingal=\ch(\xsingal)$ arrives at the receiver, it is decoded using some transformation $\rx:\iqspace\mapsto\symspace$, leading to the recovered estimate symbol $\estsym$.

\begin{figure}[t!]
\centering
\newcommand{\subfigsize}{0.22\textwidth}
\subfloat[16-QAM]{\clipbox{0cm 0cm 0cm 0cm}{\includegraphics[width=\subfigsize]{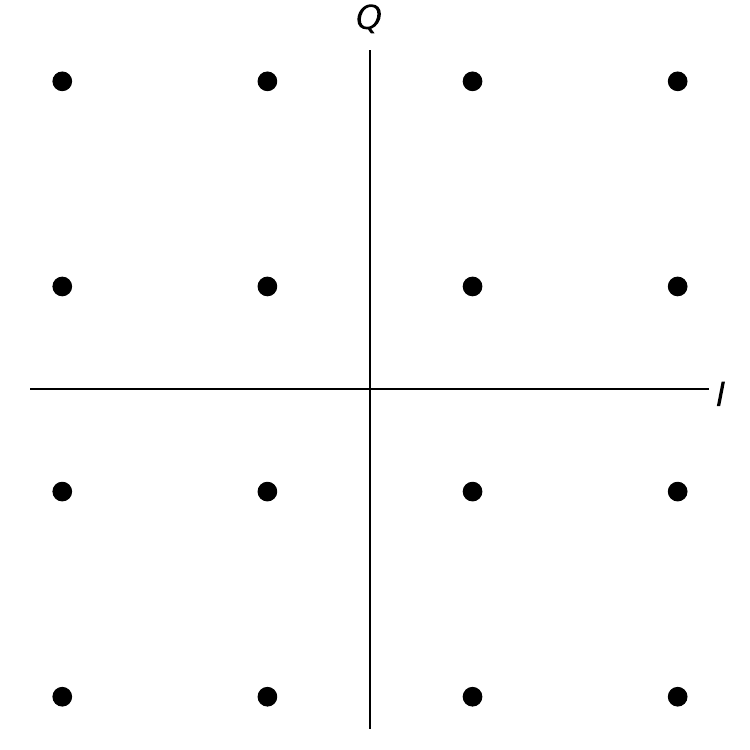}}%
\label{subfig:const_16qam}}
\hfill
\subfloat[16-PSK]{\clipbox{0cm 0cm 0cm 0cm}{\includegraphics[width=\subfigsize]{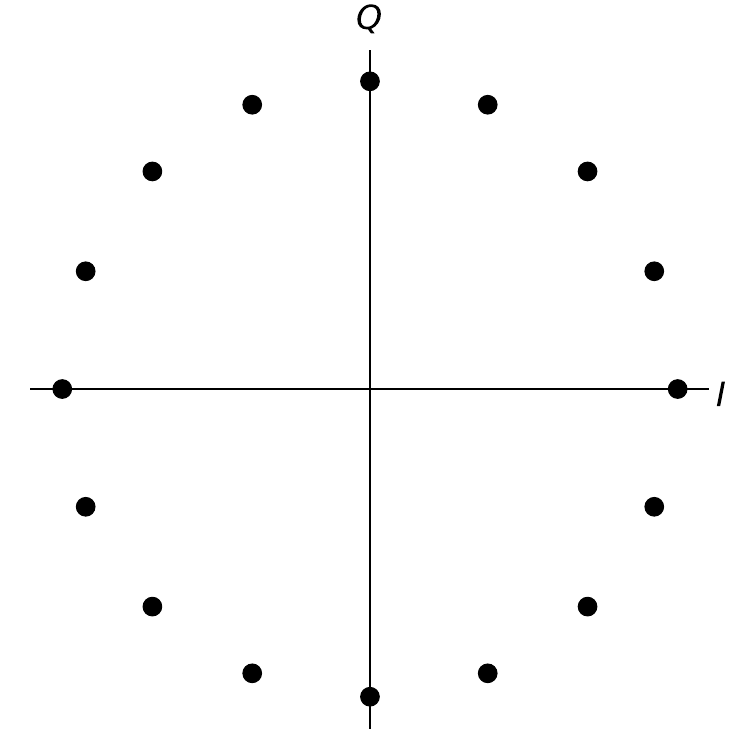}}
\label{subfig:const_16psk}}
\vspace{0.2cm}
\caption{\textbf{Examples of \glsentryshort{16qam} and 16-\glsentryshort{psk} constellation diagrams with 16 distinct symbols.}}%
\label{fig:constellation}
\end{figure}

To encode messages (symbols) into wireless transmissions (electromagnetic waves), the $\tx$-side needs to \emph{modulate} the symbols in a way that makes signal detection ($\rx$-side) as efficient as possible. In practice, this can be achieved by employing two $90^\circ$ out-of-phase carrier waves (in-phase and quadrature components) and modulating their amplitudes using \emph{amplitude-shift keying} to encode symbols (bits of information). This transformation of symbols to signals can be visualized in a 2-D plane called the I-Q plane, showing how the two components' amplitudes jointly encode a fixed number of symbols. This representation is called \textit{constellation}. An example of a \gls{16qam} constellation is illustrated in Fig.~\ref{subfig:const_16qam}.

The amount of information -- or number of bits -- a single symbol encodes depends on the size of the constellation produced by the modulation algorithm. Since the signal must adhere to certain hardware restrictions (e.g., its average power cannot exceed one), increasing the points on this constellation leads to more detection errors as certain symbols can more easily become indistinguishable due to the channel's noise. Well-known modulation schemes include \gls{bpsk} (one bit per symbol), \gls{qpsk} (two bits per symbol), and $M$-QAM a general form of \gls{qam}, where the number of bits per symbol is $k=\log_2 M$.

Other modulation techniques include \gls{psk}, which encodes information in the carrier wave's phase. Symbols on a \gls{psk} constellation diagram are placed on a unit circle with equal spacing to minimize detection errors. An example of a 16-\gls{psk} constellation is illustrated in Fig.~\ref{subfig:const_16psk}.

Degradation of signal can be measured by the ratio of erroneous to successful detection in either bits (\gls{ber}) or in symbols (\gls{ser}). 
A possible mitigation of degradation effects is using multiple channels with different characteristics (e.g., frequency diversity) to send a given message multiple times, increasing the chance of successful signal detection. 
For a given modulation scheme with $k$ bits per symbol using the channel $n$ times, the communication rate of the system is $R=k/n$. 

The most fundamental type of degradation in case of line-of-sight communication, is when symbol detection is complicated by an additive noise, that can be modelled as additive white Gaussian noise (\gls{awgn}) channel. This scenario is modeled as a probabilistic vector $\noise\in\iqspace$ sampled independently from normal distribution as $\noise\sim\mathcal{N}(0, \sigma^2\mathbf{I}_{2n})$, where the standard deviation $\sigma$ depends on the \gls{snr} and the communication rate:
\begin{equation}\label{eq:sigma}
    \sigma=\frac{1}{(2 R \cdot E_b/N_0)} \enspace, 
\end{equation}
\noindent where $E_b/N_0$ denotes the \gls{snr} per individual bit (energy per bit to noise power spectral density ratio). 
In order to model the \gls{awgn} effect, the resulting noise vector can simply be added to signal $\xsingal$ after propagation. 
In case there is no other channel effect, the received signal is modeled as:
\begin{equation}\label{eq:awgn_signal}
    \ysingal = \xsingal + \noise \enspace.
\end{equation}

A more complicated channel model, used to describe scenarios in which there is no clear line-of-sight between the transmitter and receiver is called \textit{Rayleigh channel}. This channel model represents the effect of multipath propagation, where the signal reflects off objects and creates multiple paths to the receiver~\cite{jensen2004}. 
Since the receiver detects the interfered signals of the multiple paths the received signal strength varies over time and space.
In simulations, a single-path Rayleigh fading channel is typically modeled by sampling a channel vector $\chvec\in\mathbb{R}^2$ from a normal distribution as $\chvec\sim\mathcal{N}(0, \mathbf{I}_2)$ for each transmission, with the received signal being the sum of the transmitted signal and noise. The received signal is then given by:
\begin{equation}\label{eq:rayleigh_signal}
    \ysingal = \chvec \xsingal + \noise \enspace.
\end{equation}
\noindent Note that, $\chvec$ is constant across channel uses, whereas $\noise$ is not.

\begin{table}[t!]
\renewcommand{\arraystretch}{1.5}
\begin{center}
\caption{Example Layout of a Baseline Classical Channel Autoencoder Model.}
\label{tab:classic_model}
\begin{tabular}{l|c}
\textbf{Layer}  & \textbf{Dimensions} \\
\hline
Embedding       & $M \mapsto 2n$ \\
Normalization   & $2n \mapsto 2n$ \\
\hline
Noise           & $2n \mapsto 2n$ \\
\hline
Dense + ReLU    & $2n \mapsto 4M$ \\
Dense + ReLU    & $4M \mapsto 2M$ \\
Dense + Softmax & $2M \mapsto M$
\end{tabular}
\end{center}
\end{table}

\subsection{Autoencoders in Wireless Communications}

End-to-end learning of communications systems using autoencoders arise naturally as the two problems share similarities in their structure. 
In particular, the encoder-decoder pair of an autoencoder can be regarded as parties of a communication link -- a transmitter and a receiver -- that try to \emph{encode} and \emph{decode} some message. 
In autoencoder-based communication systems (also known as channel autoencoders), the transmitter and receiver are jointly optimized to learn an arbitrary channel model. 
It has been shown to be a very versatile model which can be augmented to handle more interesting channel models~\cite{7886039,8054694,dorner2017deep,o2017deep,zou2021channel,GARCIA2022192,9446711,9204706,9154335}.

The general description of the Channel Autoencoder model is as follows. 
First, the encoder implements the transmitter-side function $\tx$ to create a mapping from each symbol to some signal (latent space constellation). 
This mapping often resembles \gls{qam} or \gls{psk} constellations. 
However, since there is no predefined mathematical channel model, any valid mapping is possible as long as it minimizes communication error. 
An encoder can be defined in many ways. It can be implemented as a NN with multiple dense or convolutional layers, or even as simple embedding lookup table. 

Next, the channel introduces some probabilistic variations into the latent space values via function $\ch$. 
These variations can be, for example, approximations of real channel models. 
Previous implementations of channel simulations for autoencoders include the effects of \gls{awgn} noise, Rayleigh channels, phase shift, other hardware impairments, and even \gls{mimo} channels~\cite{bourtsoulatze2019deep}. 
Finally, the decoder implements receiver-side function $\rx$, and learns to distinguish distinct code points of the latent space corrupted by the channel impairment. 
Possible implementations include NN of dense layers with Softmax activation as final layer. 
More robust models also utilize radio transformer networks (RTNs)~\cite{8054694} that use a separate stack of dense layers to learn the inverse of the channel effect (e.g., fading). 

The channel autoencoder's input is the symbol $\sym$, which can be directly fed to an embedding lookup table-based encoder or converted into a one-hot vector in the case of stacked dense layers. 
The model's output is a probability vector of size $M$ to classify the decoded message. 
Learning is usually done using a gradient descent-based technique, with (sparse) categorical cross entropy as the loss function. 
Table~\ref{tab:classic_model} illustrates a baseline classical Channel Autoencoder architecture, and Ref.~\cite{zou2021channel} provides an overview of state-of-the-art implementations. This approach of using autoencoders in wireless communications has shown promising results in various scenarios, paving the way for more advanced and robust communication systems.

\section{Hybrid Channel Autoencoders}\label{sec:qpae}

\begin{figure}[t!]
\centering
\includegraphics[height=0.3\textwidth]{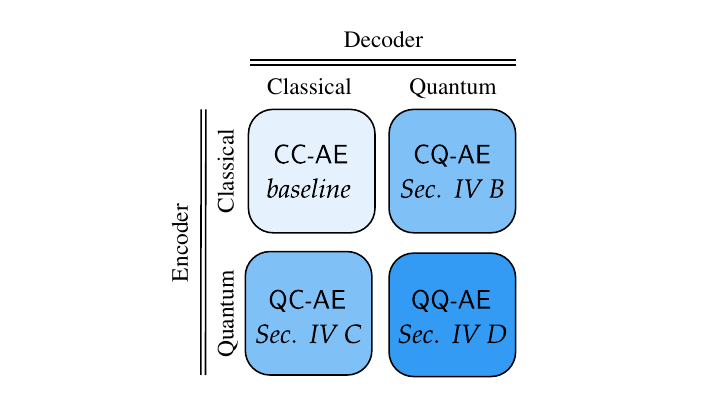}
\caption{\textbf{Possible variants of the Hybrid Channel Autoencoder ($\hybrid$).} The models range from a purely classical approach to mixed models, where either the transmitter, the receiver, or both are implemented as \glspl{qnn}. The most advanced variant utilizes \glspl{qnn} for both the transmitter and receiver, potentially maximizing the benefits of quantum computation in communication systems. Each configuration offers unique advantages and trade-offs in terms of computational complexity, adaptability to different channel conditions, and potential for performance improvements.}
\label{fig:variations}
\end{figure}

Quantum autoencoders are analogous to classical autoencoders, as they were first introduced to compress and denoise quantum data~\cite{Romero_2017, PhysRevLett.124.130502}.
Furthermore, in Ref.~\cite{PhysRevResearch.1.033063}, the authors proposed a hybrid scheme with a classical encoder and a quantum decoder.
Building on these results, we take a slightly different approach and introduce a hybrid quantum-classical scheme for the Channel Autoencoder, in which both the encoder and the decoder can be either classical or quantum as shown in Fig.~\ref{fig:variations}.
Since the channel in this context is always classical, the latent space is also classical, meaning that the bottleneck of our model involves measurement and classical data encoding at the $\tx$ and $\rx$ side for a quantum \emph{encoder} and \emph{decoder}, respectively.
Using both classical and quantum components enables the construction of a flexible hybrid pipeline.
The rest of this section explores possible implementations of different parts of the system based on \gls{qnn} input and output requirements.

We investigate the practical implications of these choices for the Hybrid Channel Autoencoder ($\hybrid$) model in Sec.~\ref{sec:experiments} as part of hyperparameter search with performance measurement.

\begin{figure}[htpb]
\centering
\includegraphics[width=0.25\textwidth]{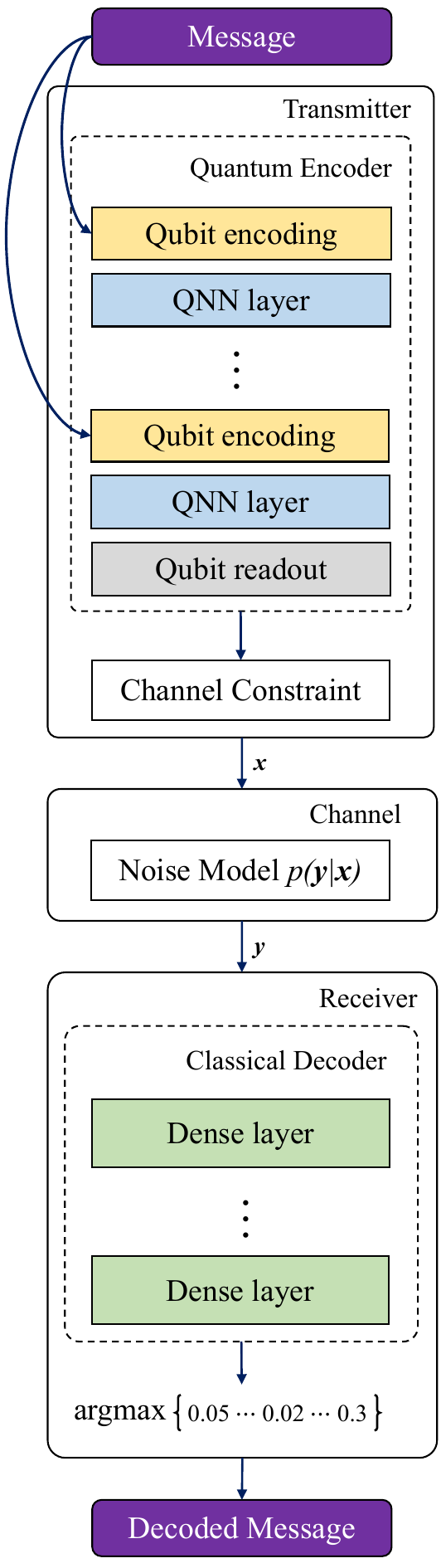}
\caption{\textbf{End-to-end Communication via a Quantum Autoencoder.} This figure depicts a hybrid Channel Autoencoder with a \gls{qnn} transmitter and a classical NN receiver. The \gls{qnn} leverages data re-uploading for accelerated learning. The the channel is characterized by a conditional probability representing message degradation.}
\label{fig:quantum_tx}
\end{figure}

\begin{figure*}[htbp]
\captionsetup[subfloat]{captionskip=10pt}
\centering
\newcommand{\subfigsize}{0.25\textwidth}
\subfloat[$\QC{1}$ encoder]{\includegraphics[height=2.5cm]{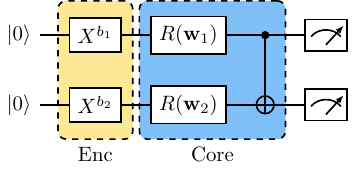}%
\label{subfig:circ_encoder_1}}
\hspace{0.3cm}
\subfloat[$\QC{2}$ encoder]{\includegraphics[height=2.5cm]{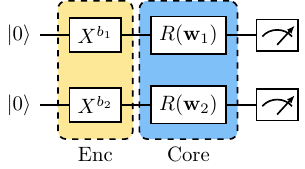}%
\label{subfig:circ_encoder_2}}
\hspace{0.3cm}
\subfloat[$\CQ{1}$ decoder]{\includegraphics[height=2.5cm]{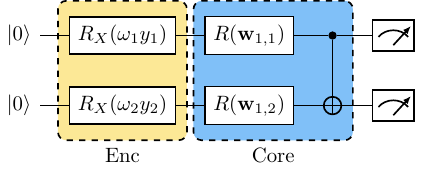}%
\label{subfig:circ_decoder_1}}

\vspace{0.2cm}
\subfloat[$\CQ{2}$ decoder]{\includegraphics[height=2.5cm]{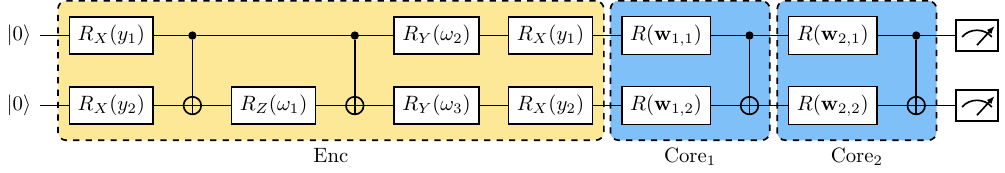}%
\label{subfig:circ_decoder_2}}

\vspace{0.2cm}
\caption{\textbf{Best performing quantum encoder and decoder implementations for \glsentryshort{4qam} modulation.} 
Each Ansatz circuit consists of an encoding block, a core layer, and a measurement in the computational basis. 
Subfigures (a) and (b) display quantum encoder circuits using basis embedding with a single core layer composed of single-qubit rotation gates. Furthermore, the core layer in (a) also includes an additional $CNOT$. 
Subfigure (c) features a quantum decoder with weighted angle embedding and a core layer identical to that in (a). 
Subfigure (d) introduces another quantum decoder that employs a more complex time evolution embedding inspired by \glsentryshort{qaoa} circuits. 
The core layer is the same as in (a) and (c), but it is repeated twice.}
\label{fig:ansatze}
\end{figure*}

\subsection{Hybrid Model Variations}\label{sec:variations}

In this section, we construct the Quantum $\tx$, Quantum $\rx$, and Quantum $\tx$-$\rx$ $\hybrid$ models by outlining their core elements. These elements are versatile enough to support widely-used schemes, including \gls{psk} and \gls{qam} modulations. While our initial focus is on a single channel use with an emphasis on scaling in $M$, we also describe the generalization of this methodology to accommodate multiple channel uses. We begin with the \gls{awgn} channel of Eq.~\eqref{eq:sigma} as a foundational model and subsequently adapt the $\hybrid$ to address more intricate channel impairments when $M=4$.

For every model, we introduce \textit{feature encoding} strategies to efficiently route the classical input into the \gls{qnn}. We then define the \textit{core layers} of the \gls{qnn}, which can be iteratively used (\textit{layer number}) to increase the network's parameter count. These layers, available in various configurations, are tailored for distinct tasks or scenarios. An optional \textit{data re-uploading} technique~\cite{PerezSalinas2020DataRF} can be employed within core layers to boost the network's expressivity. 
Finally, we establish a \textit{measurement} procedure compatible with the \gls{qnn}'s output specifications. 
In the simplest case, only the core layer uses trainable parameters, but we can also integrate such weights into the feature encoding and measurement operators.

Fig.~\ref{fig:quantum_tx} provides a visual representation of our Quantum $\tx$ model, showcasing the foundational structure of our hybrid models. This illustration offers insight into the layout and functionality of our hybrid approach. 
The Quantum $\rx$ model can be constructed analogously, as presented in our previous work~\cite{tzs20222hybrid}. Consequently, the Quantum $\tx$-$\rx$ model is simply the combination of these two. 

\subsection{Quantum TX model}\label{sec:4qamtx}

To realize the $\tx$ model, we start by encoding the input features from $\symspace$. A direct approach involves basis encoding into a $\log_2{M}$-qubit system. This is because representing $M$ symbols in binary requires $\log_2{M}$ bits, keeping in mind that in modulation schemes, $M$ is invariably a power of two. Formally, the unitary for this encoding strategy for a symbol $\sym \in \symspace$ is expressed as: 
\begin{equation}\label{basis4qam}
    U_{enc1}(s) = \prod\limits_{i=1}^{\log_2{M}} X_i^{b_i} \enspace ,
\end{equation}
\noindent where $b_i$ denotes the $i$-th bit of the binary representation of the number $s-1$.

Another approach is using a discretized version of angle encoding. For a single channel use, we can apply single qubit rotations $ \alpha, \beta \in \{X, Y\}$ on two qubits to get: 
\begin{equation}
    U_{enc2}(s) = R_{\alpha}\left(\frac{\pi}{M}s\right) \otimes R_{\beta}\left(\frac{\pi}{M}s\right) \enspace ,
    \label{eq:disc_ang}
\end{equation}
\noindent or using multiple rotations we arrive at: 
\begin{equation}
\begin{split}
    U_{enc3}(s) & = R_{Z}\left(\frac{\pi}{M}s\right) R_{\alpha}\left(\frac{\pi}{M}s\right)\\
    & \otimes R_{Z}\left(\frac{\pi}{M}s\right) R_{\beta}\left(\frac{\pi}{M}s\right) \enspace.
\end{split}
\end{equation}

Furthermore, we can utilize \emph{trainable weights} so that the \gls{qnn} automatically learns how to convert symbols to rotation angles as: 
\begin{equation}
    U_{enc4}(\mathbf{w},s) = R_{\alpha}(w_1 s) \otimes R_{\beta}(w_2 s) \enspace,
    \label{eq:weight_ang}
\end{equation}
\noindent or as: 
\begin{equation}
\begin{split}
    U_{enc5}(\mathbf{w},s) & = R_{Z}\left(w_1 s\right) R_{\alpha}\left(w_2 s\right)\\
    & \otimes R_{Z}\left(w_3 s\right) R_{\beta}\left(w_4 s\right) \enspace,
\end{split}
\end{equation}
where $\mathbf{w}$ is a vector of trainable weights that are initialized randomly.

The second part of our quantum circuit, the core layer, represents the primary operation of the \gls{qnn} and always incorporates trainable weights. In general, a core layer is formulated as:
\begin{equation}\label{eq:core}
U_{core1}(\mathbf{W}) = \prod\limits_{i,j\in K} C_i(X_j)\prod\limits_{i=1}^{k} G_i(w_{i,1}, w_{i,2}, w_{i,3}) \enspace,
\end{equation}
where $k$ denotes the number of qubits, $K$ is a configurable ordered list of qubit pairs and $\mathbf{W}$ is a matrix of trainable weights initialized randomly.
In the above equation, entanglement can be tailored by carefully choosing elements of $K$. For instance, we can employ a strongly entangling layer introduced in Ref.~\cite{PhysRevA.101.032308}, known to be highly expressive.
Alternatively, we might opt for core layers with limited entanglement to circumvent barren plateaus.

The core layer's repetition can enhance both the parameter count and the quantum circuit's expressivity. This repetition frequency is referred to as the \emph{layer number}. When layers are repeated, there is an option to incorporate data re-uploading, especially with angle or basis encoding. Given that the output of $\tx$ for $n=1$ should be a vector in $\mathbb{R}^2$, we should define a measurement method yielding two real values. For a $2$-qubit system, a logical approach is to consider the expected value of a local observable on each qubit separately.
This involves executing projective measurements and averaging the results over a reasonable number of shots:
\begin{equation}\label{eq:measurement1}
    output_1 = \left(\left\langle \sigma_1^Z \right\rangle , \left\langle \sigma_2^Z \right\rangle \right) \enspace.
\end{equation}

Combining these foundational elements, we formulate our first hybrid model, the $\qcae$, depicted in Fig.~\ref{fig:quantum_tx}.
Furthermore, for $M=4$, any combination of encoding, core layer, and measurement can be employed on a $2$-qubit system without additional adjustments.

However, adapting the model to higher values of $M$ requires adjustments of some of the building blocks. Specifically, with basis encoding, the qubit count scales logarithmically with $M$. This increase in the number of qubits also necessitates a new choice of measurement observables. For instance, with $M=16$ on a $4$-qubit system, measurement can be defined as:
\begin{equation}
    output_2 = \left(\left\langle \sigma_1^Z \otimes \sigma_2^X \right\rangle , \left\langle \sigma_3^Z \otimes \sigma_4^X \right\rangle \right) \enspace,
\end{equation}
or we can perform the measurement of the previous type on a subset of the qubits.

In contrast to  basis encoding, discretized angle encoding remains functional as is with just two qubits, regardless of the choice of $M$.
This characteristic makes discretized angle encoding a favorable choice for preliminary experimental evaluation, as increasing the qubit number can lead to prohibitive simulation scenarios. 

\subsection{Quantum RX model}\label{sec:4qamrx}

For a single channel use ($n=1$), the input to the decoder side is always a vector of $\mathbb{R}^2$ representing the corrupted constellation point. Given that the output of the decoder must be a probability vector of size $M$, we propose using a qubit system of size $\log_2{M}$. We can encode two real numbers in a $\log_2{M}$-qubit system in various ways.

We focus on time evolution encoding methods due to their flexibility. The first method we consider is angle embedding with single-qubit rotations. For $M=4$, this encoding can be directly achieved by associating a distinct qubit to both features with optional weights:
\begin{equation}
U_{enc_6}(\mathbf{w}, \mathbf{y}) = R_{\alpha}\left(w_1 y_1\right) \otimes R_{\beta}\left(w_2 y_2\right) \enspace,
\label{eq:enc6}
\end{equation}
\noindent where $\mathbf{y}$ represents the channel output and $\mathbf{w}$ is a vector of weights. It is worth noting that by setting weights to a constant $\mathbf{1}$, trainable weights can be removed.

For cases where $M>4$, we have two alternatives: either use only the first two qubits for embedding or encode $\mathbf{y}$ across multiple subsystems simultaneously. The latter, \textit{parallel encoding} is known to substantially increase the expressivity, as shown in Ref.~\cite{tzs20222hybrid}, and can be defined as:
\begin{equation}
\begin{split}
    U_{enc_7}(\mathbf{w}, \mathbf{y}) & = R_{\alpha}\left(w_1 y_1\right) \otimes R_{\beta}\left(w_2 y_2\right) \\
    & \otimes R_{\alpha}\left(w_3 y_1\right) \otimes R_{\beta}\left(w_4 y_2\right),\, \mathrm{if} \, M=16 \enspace.
\end{split}
\end{equation}
\noindent In the above equation, setting $w_3 = w_4 = 0$ effectively disables parallel encoding, reverting to the first option.

We also explored another Hamiltonian $H$ for time evolution encoding, drawing inspiration from the \gls{qaoa} Ansatz~\cite{2014arXiv1411.4028F}. In our numerical experiments we used the \textit{Pennylane} software library's implementation of this approach~\cite{PennylanePaper}, which can be seen as applying angle embedding twice interposed, with $ZZ$ gates inserted in between, and these gates having trainable weights. 

Given the inherent complexity of decoding compared to encoding, we anticipate the necessity for a higher number of layers. Consequently, for the core layers, we can employ Eq.~\eqref{eq:core}, which allows for multiple layers and the option of data re-uploading. 

Measurements can be performed in the basis of any local observable across all qubits. The resulting probability distribution over the $M$ basis states serves as the decoder's output:
\begin{equation}
output_3 = \{P_{\mathbf{\theta}}(b)\;|\; b = (s-1)_2, s \in \symspace \} \enspace,
\label{eq:output3}
\end{equation}
\noindent where, according to the Born-rule,
\begin{equation}
P_{\mathbf{\theta}}(b) = \left|\langle b | \psi(\mathbf{\theta})\rangle\right|^2 \enspace,
\end{equation}
\noindent and $(a)_2$ denotes the binary representation of non-negative integer $a$. This equation yields the final output of the quantum circuit, from which we can directly determine the most likely transmitted symbol from $\symspace$.

\subsection{Quantum TX-RX model}
We can combine the two previous models ($\qcae$, $\cqae$) while keeping the classical channel to arrive at a solution with quantum models at both ends. 
The $\qqae$ uses a \gls{qnn} to learn $\tx$, transmits signals through a classical channel to be decoded by another \gls{qnn} implementing $\rx$. 
Since the two \glspl{qnn} work separately, we can utilize separate designs for each, as long as the input and output of these systems are compatible. 

\subsection{Model Generalization}

The $\hybrid$ model is already built with modulation scaling in mind, capable of handling $M$-QAM modulations of any $M$ that is a power of two. 
However, it's worth noting that increasing the modulation complexity would necessitate larger models, quickly leading to computational complexities that are prohibitive for classical simulation.
Another direction toward scaling our model involves enhancing its capability to work on more realistic channel impairments.
A straightforward way of achieving this is to increase the number of channel uses so that there is more redundancy in the system. 
This way, we will potentially be able to adapt our model to a more challenging channel scenario, namely Rayleigh fading, defined in Eq.~\eqref{eq:rayleigh_signal}. 
To demonstrate this, we construct a Quantum $\rx$ hybrid model with a \gls{4qam} modulation scheme in mind, generalized for two channel uses and optimized for Rayleigh fading.
It is worth noting that the general approach for increasing channel uses can be similarly applied to arbitrary noise conditions.

In order to increase the number of channel uses $n$, we need to allow larger input vectors.
For $n=2$, the decoder model's input dimension must be scaled to $\mathbb{R}^4$. This can be achieved by extending the angle encoding, as described in Eq.~\eqref{eq:enc6}, to $4$ inputs and $4$ qubits. 

The core layer in Eq.~\eqref{eq:core} can already handle larger inputs with an adequate $k$. 
For this particular model, we set $k=4$. 
Additionally, we can consider core layers that use rotations around a specific axis, such as only $R_Y$ rotations, instead of general rotation gates.

The decoder's output is the probability distribution, as defined in Eq.~\eqref{eq:output3}. However, since we have more basis states than symbols, we measure only a subset of the qubits --- for example, the first two in a $4$-qubit system.

\begin{figure*}[htbp]
\centering
\includegraphics[width=0.95\textwidth]{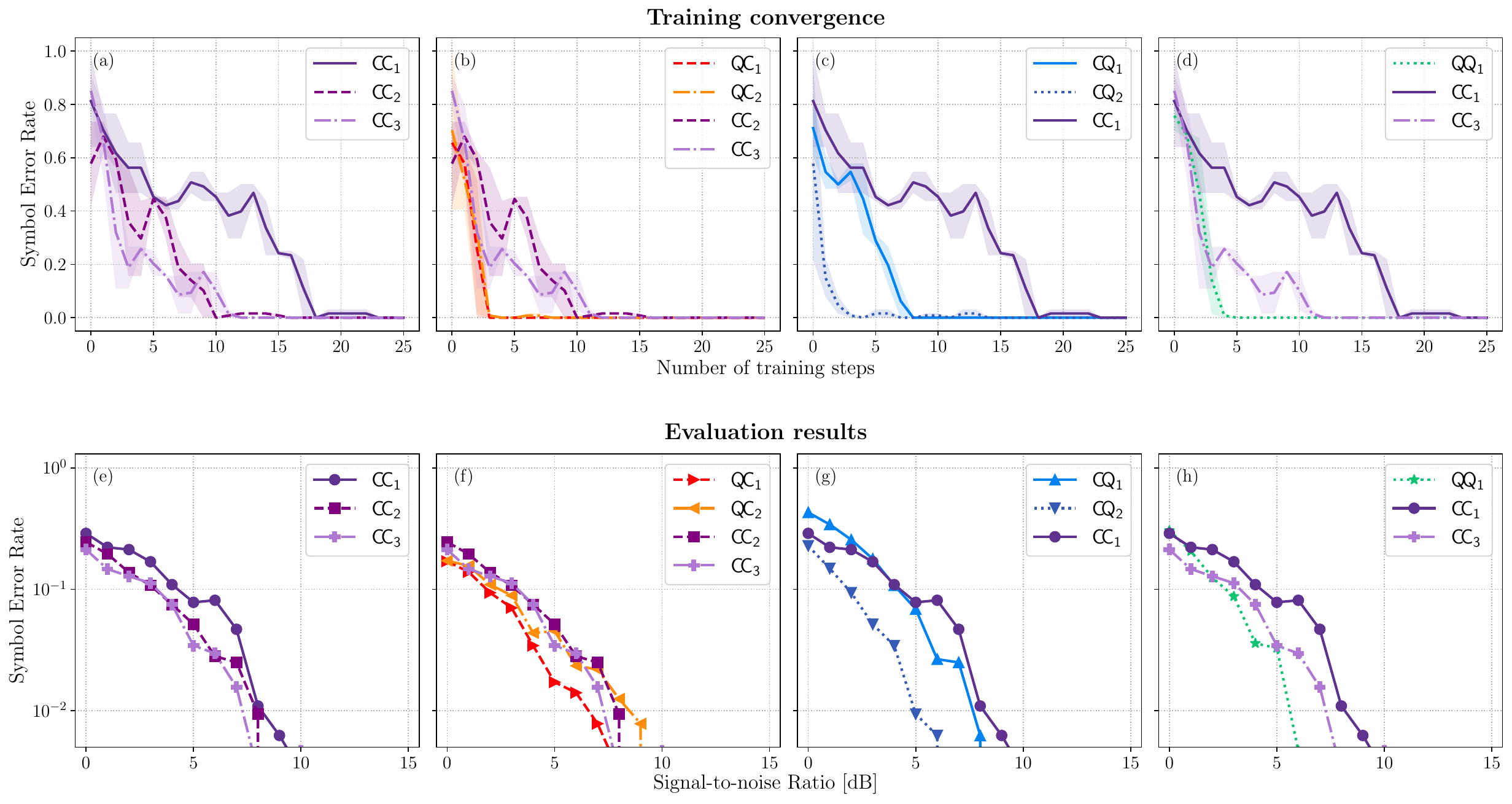}
\caption{\textbf{Training convergence (top row) and evaluation results (bottom row) for \glsentryshort{4qam} with an \glsentryshort{awgn} channel.} The figure compares the performance of top-performing hybrid models against classical baselines, all trained at a $15\;\text{dB}$ \gls{snr}. We observe faster training for the quantum-enhanced models (b)-(d), while also improving generalization on noise levels not used for training (f)-(h). (a)-(d) illustrate convergence metrics, while (e)-(h) focus on generalization. (a) and (e) set the performance baseline by comparing purely classical autoencoders, which is used for subsequent comparisons with hybrid models having similar parameter counts. (b), (c), and (d) examine the convergence of $\qcae$, $\cqae$, and $\qqae$ models, respectively. Correspondingly, (f), (g), and (h) present their generalization performance using \gls{ser} evaluated at increasing \gls{snr} levels.}  
\label{fig:4qam_results}
\end{figure*}

\begin{table}[htbp]
\renewcommand{\arraystretch}{1.5}
    \centering
    \begin{tabular}{|c||c|c||c|c||c|}
        \hline
        Model & TX    & TX        & RX    & RX        & Total     \\[-0.15cm]
        name  & model & params & model & params & params \\
        \hline\hline
        $\CC{1}$ & lookup & 8 & dense & 24 & 32 \\
        $\CC{2}$ & lookup & 8 & dense & 220 & 228 \\
        $\CC{3}$ & dense & 23 & dense & 220 & 243 \\
        \hline
        $\QC{1}$ & quantum & 6 & dense & 220 & 226 \\
        $\QC{2}$ & quantum & 6 & dense & 220 & 226 \\
        \hline
        $\CQ{1}$ & lookup & 8 & quantum & 8 & 16 \\
        $\CQ{2}$ & lookup & 8 & quantum & 15 & 23 \\
        \hline
        $\QQ{1}$ & quantum & 6 & quantum & 8 & 14 \\
        \hline
    \end{tabular}
    \caption{\textbf{Parameter counts for classical and hybrid quantum channel autoencoder models.} This table lists the number of parameters for each model, considering both lookup table-based and dense-layered configurations, as used in the \gls{4qam} \gls{awgn} experiments.}
    \label{tab:4qam_models}
\end{table}

\section{Numerical Experiments for Performance Comparisons}\label{sec:experiments}
To present a comparative analysis, we designed a series of simulated numerical experiments, that aim to compare the hybrid models against fully classical baselines.
Quantum circuit simulations and \glspl{qnn} training are carried out using the \textit{Pennylane} software package~\cite{Bergholm2018PennyLaneAD} with its Tensorflow~\cite{tensorflow2015-whitepaper} interface. 
We employ the PennyLane-Lightning Plugin to speed up the simulation and differentiation of the quantum circuits. 
For optimization, we utilize the Adam algorithm~\cite{kingma2014adam}, selecting learning rates from the set $\{0.1, 0.01, 0.001\}$ as a part of the hyperparameter search.

The training batch size was set to $64$ (determined empirically) to allow the simulation of larger circuits without significantly impeding the training process.
During training, we fix the \gls{snr} at $15\;\text{dB}$ but also validate the results under various noise levels.

\subsection{Model Selection Methodology}

The first step of model selection is the definition of relevant benchmark metrics.
We measure the accuracy using the Symbol Error Rate (\gls{ser}), a commonly used key performance indicator in radio communication.
During training, we initially concentrate on achieving the lowest possible \gls{ser} value and the rate at which the models converge to this value. 
These two factors establish the baseline for our selection procedure. To further validate our approach, we assess the generalization capabilities of the trained autoencoders by testing them on unseen \gls{snr} levels, conducting ten-batch tests for each level. 
Additionally, we aim to minimize the number of parameters while still maintaining strong performance. 
For quantum encoders and decoders, we also take into account the circuit depth, a critical cost metric for near-term quantum devices.

To benchmark the hybrid models, we establish fully classical baselines, taking into account the aforementioned factors.
We considered two kinds of classical encoders: the first one being a linear lookup table with $2 n M$ parameters and the second one being a feedforward neural network with one fully connected hidden layer of tunable size. 
The classical decoder consists of two fully connected hidden layers, both with adjustable sizes.
The classical baseline models are designed to have a parameter count that is comparable to their quantum counterparts.

We use KerasTuner~\cite{omalley2019kerastuner} to explore the hyperparameter space, which is constructed by combining the elements of $\hybrid$ as outlined in Sec.~\ref{sec:qpae}. In addition to these base elements, we also consider other tunable hyperparameters like the number of layers, learning rate, and the option for data re-uploading.
Due to the prohibitive size of this search space, we apply empirical pre-filtering based on initial results for each specific model.
When tuning the hybrid models $\cqae$ and $\qcae$, we fix the classical encoder and decoder to previously tested, high-performing models.

\subsection{4-QAM Simulations over AWGN Channel}

To find the optimal balance between performance and model complexity, we conducted an extensive hyperparameter search using \gls{4qam} modulation. 
The resulting empirically selected models and their parameters are listed in Table~\ref{tab:4qam_models}. 
Figure~\ref{fig:4qam_results} illustrates the simulation results, showcasing the models' convergence properties and evaluation performance. 
For easier comparison, models are grouped together in sub-plots to highlight the performance differences between quantum and classical approaches, as well as among various classical configurations. 

Among several classical baselines (Fig.~\ref{fig:4qam_results} (a) and (e)), we first select a model with lookup table-based transmitter that performs relatively well while minimizing the number of parameters in the receiver ($\CC{1}$). 
The second classical model also consists of a lookup table at the transmitter, but the receiver is replaced with a much larger NN ($\CC{2}$). 
Our third choice uses a dense-layered transmitter with comparable performance to the previous ones for the price of a slight increase in parameter count ($\CC{3}$). 

First, we benchmark quantum encoders. For this, we showcase two simple Ans{\"a}tze using basis embedding and core layers composed of general rotation gates. 
We consider the core layer with and without entanglement, denoted as $\QC{1}$ and $\QC{2}$, respectively. 
The corresponding quantum circuits are illustrated in Fig.~\ref{subfig:circ_encoder_1} and Fig.~\ref{subfig:circ_encoder_2}. 
With just a single layer, the model not only reduces the number of parameters but also achieves better \gls{ser} convergence compared to any classical baseline of a similar size. 
Furthermore, the model employing a quantum transmitter with entanglement exhibits slightly better generalization.
The corresponding results are shown in Fig.~\ref{fig:4qam_results} (b) and (f).

Regarding quantum decoders, our model significantly reduces the number of parameters, down to $8$, compared to the smallest classical model, which has $32$ parameters.
With a single-layered Ansatz with weighted angle embedding, shown in Fig.~\ref{subfig:circ_decoder_1}, the $\CQ{1}$ model can achieve faster convergence and comparable generalization with only $8$ tunable parameters (subfigures (c) and (g) of Fig.~\ref{fig:4qam_results}). 
To demonstrate the benefits of using more layers and a different Hamiltonian for time evolution encoding (see Fig.~\ref{subfig:circ_decoder_2}), we examine the performance of the $\CQ{2}$ model, that has two layers with \gls{qaoa} embedding.
This approach indeed accelerates learning and improves generalization, although at the cost of a deeper circuit and a higher parameter count.

Lastly, we evaluate the $\qqae$ model, which optimally combines the encoder from $\QC{1}$ and the decoder from $\CQ{1}$.
This configuration minimizes the total number of tunable parameters to $14$ and also enhances generalization performance (Fig.~\ref{fig:4qam_results} (d) and (h)).
These numerical results clearly show that the more components of the end-to-end architectures are adapted for \glspl{qnn} solution, the more we can reduce the number of trainable parameters, train faster for same accuracy level, all the while either keeping or even enhancing generalization capabilities of the trained models on an extended range of noise levels.

\begin{figure}[t!]
\centering
\includegraphics[width=0.48\textwidth]{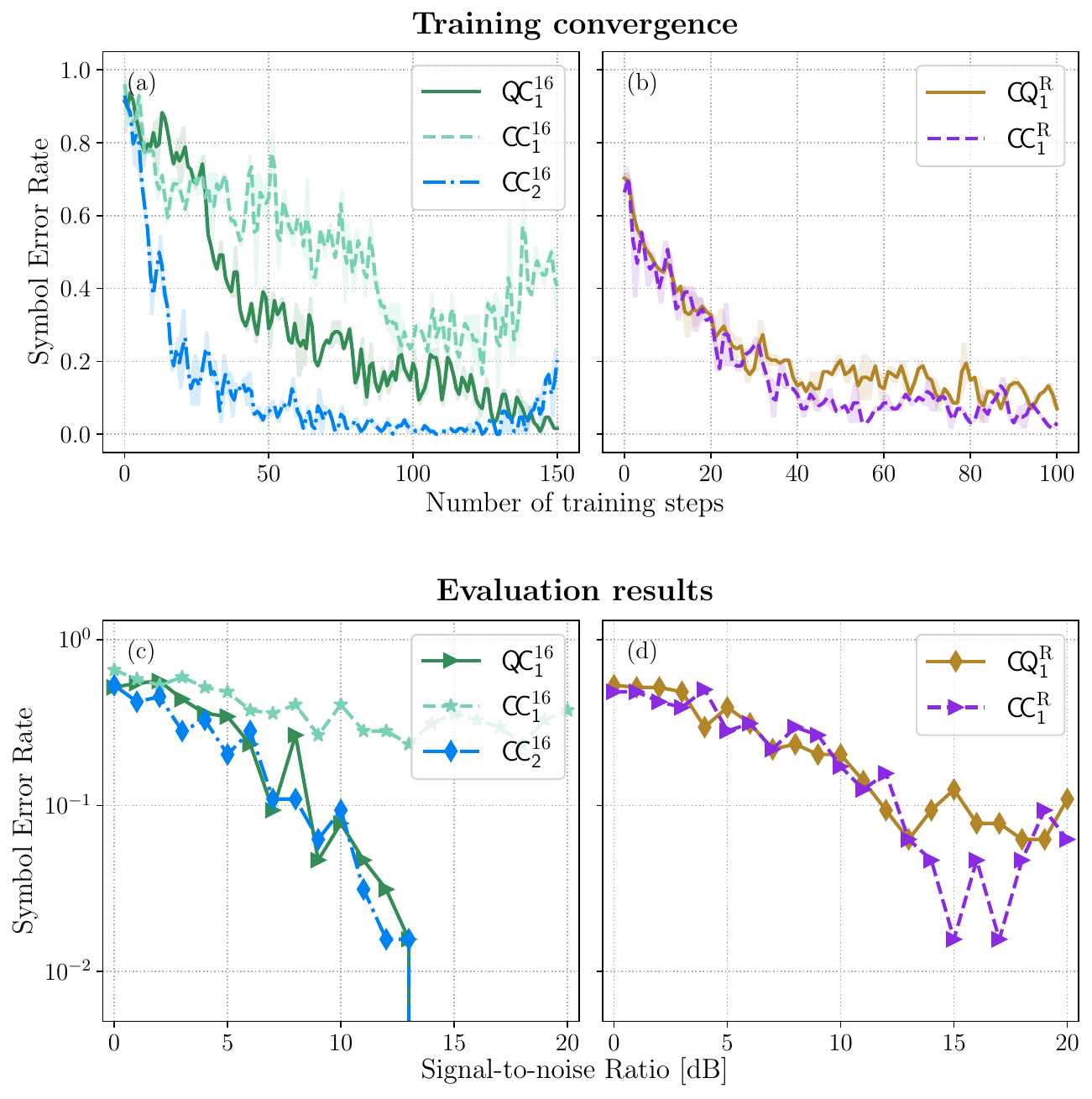}
\caption{\textbf{Training convergence (top row) and evaluation results (bottom row) for \glsentryshort{16qam} over \glsentryshort{awgn} and \glsentryshort{4qam} with Rayleigh fading channels.} Comparing two specific $\hybrid$ models with their respective classical baselines, we observe quantum-enhanced architectures reducing the number of trainable parameters without loss of performance. (a) and (c) present the convergence and evaluation metrics for the $\qcae$ model, which uses \gls{16qam} modulation over an \gls{awgn} channel. (b) and (d) illustrate the performance of the $\cqae$ model, which employs \gls{4qam} modulation over a Rayleigh fading channel with two channel uses.}
\label{fig:ray_plus_16qam_results}
\end{figure}

\subsection{16-QAM Simulations over AWGN Channel}

We demonstrate the utility of the weighted angle embedding method for discrete values, introduced in Eq.~\eqref{eq:weight_ang}, through a series of experiments using \gls{16qam} modulation and AWGN noise.
We implement a $\qcae$ model with a quantum encoder circuit on $2$ qubits, featuring an Ansatz composed of two core layers and weighted data re-uploading. This encoder has $14$ parameters.
We compare this model against a $\ccae$ model that employs a fully-connected classical neural network for the transmitter. 
Utilizing hyperparameter search to find the optimal balance between parameter count and model performance, we find that the smallest classical model achieving comparable performance to our $\qcae$ has significantly more parameters in the transmitter --- $306$ for the $\CC{2}^{16}$ model.
We also introduce another classical model, $\CC{1}^{16}$, which has a dense-layered transmitter restricted to only $40$ parameters. This is done to emphasize that classical models require a substantially larger number of parameters as the complexity of the problem increases. This becomes evident when the $14$-parameter $\QC{1}^{16}$ model already yields superior results.
The comparison of these three models is shown in Fig.~\ref{fig:ray_plus_16qam_results} (a) and (c).

Our results indicate that a system with only $2$ qubits can effectively solve the channel autoencoder problem for higher values of $M$. While basis embedding would require $\log M$ qubits, this method requires exactly $2$ qubits for one channel use.
For an extensive study on \gls{16qam} $\cqae$ models, we refer the reader to the results in Ref.~\cite{tzs20222hybrid}.

\subsection{4-QAM Simulations with Rayleigh Fading}

We extend our model variations by introducing more challenging channel impairments for which a single channel use is insufficient. 
Specifically, we implement a $\cqae$ model with \gls{4qam} modulation and apply Rayleigh fading to the channel to demonstrate how our $\hybrid$ framework can incorporate two channel uses. 
This model employs weighted angle embedding and core layers consisting solely of $R_Y$ rotations and $CNOT$ gates.

To handle these more complex noise scenarios, we introduce a series of additional refinements to the basic quantum decoder model structures.
First, we discover that adding a pre-processing layer to the decoder encoding step, a simple transformation of the channel output, e.g., $\arctan(\mathbf{y})$, improves performance.
Second, we introduce the use of ``measurement weights,'' i.e, additional parametrized single-qubit rotation gates as a last layer before the final measurements. This technique effectively improves convergence rates and model generalization. 

After extensive hyperparameter tuning, our final model $\CQ{1}^{R}$ employs $16$ layers and $132$ parameters on the receiver side. This model achieves performance comparable to our chosen classical baseline, which has a similar parameter count of $150$. The results are presented in subfigures (b) and (d) of Fig.~\ref{fig:ray_plus_16qam_results}, where we observe quantum-enhanced architectures reducing the number of trainable parameters without loss of performance.

\section{Conclusion}\label{sec:summary}

A comprehensive study was presented on the design and evaluation of hybrid quantum-classical channel autoencoder architectures for radio communication. In this $\hybrid$ framework we demonstrated the adaptability of the model to various modulation schemes and channel conditions including \gls{4qam} and \gls{16qam} modulations, as well as the damaging effects of \gls{awgn} and Rayleigh fading.
The presented novel hybrid and fully quantum channel autoencoder designs achieved comparable and in some cases superior performance to classical baselines, and did so with fewer trainable parameters. The most significant performance enhancement was reached with the introduction of quantum encoders, with significant reduction in trainable parameters.

We leave for future work optimizing such framework for adapting to near-term quantum devices, for scaling the models, especially as the modulation complexity increases, and for exploring different encoding schemes against various types of harsh radio environments in real-world scenarios.

\section*{Acknowledgment}
The authors would like to thank the support of the Hungarian National Research, Development and Innovation Office (NKFIH) through the KDP-2021 and KDP-2023 funding scheme, the Quantum Information National Laboratory of Hungary and the grants TKP-2021-NVA-04 and FK 135220.

\onecolumn
\newpage
\twocolumn

\bibliographystyle{IEEEtran}       %
\bibliography{IEEEabrv,references} %

\onecolumn
\newpage
\twocolumn

\begin{IEEEbiography}[{\includegraphics[width=1in,height=1.25in,clip,keepaspectratio]{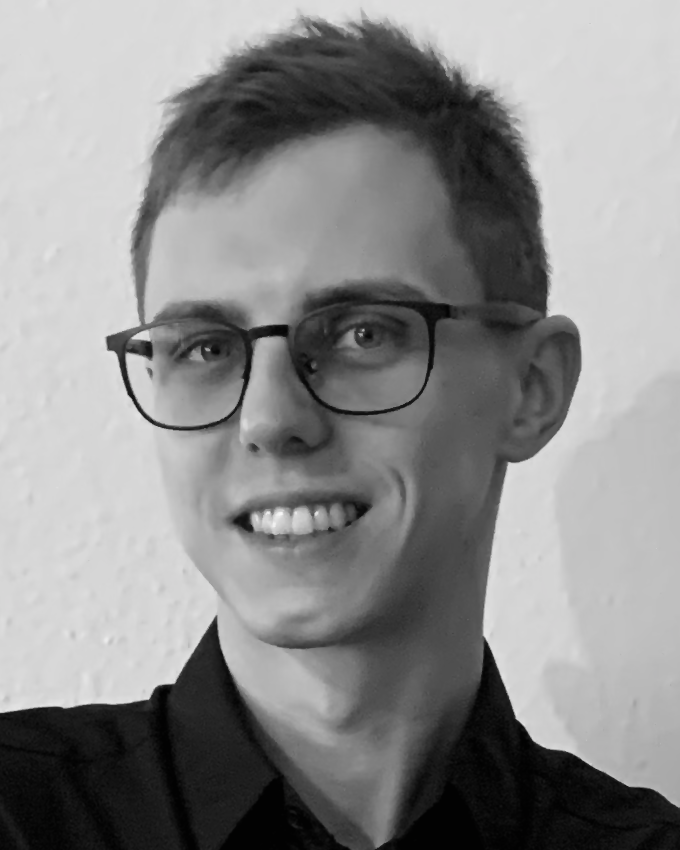}}]{Zsolt I. Tabi} was born in Budapest, Hungary in 1990. 
He received his B.Sc. and M.Sc. degrees in computer science from E\"otv\"os Lor\'and University, Budapest, Hungary, in 2016 and 2018, respectively. 
He is currently pursuing a Ph.D. degree in computer science at E\"otv\"os Lor\'and University and is collaborating with the Ericsson Research Team. 
He has been employed as a software developer since 2016. 
He also teaches programming languages at E\"otv\"os Lor\'and University. 
His research interests include programming language design and quantum computing. 
\end{IEEEbiography}

\begin{IEEEbiography}[{\includegraphics[width=1in,height=1.25in,clip,keepaspectratio]{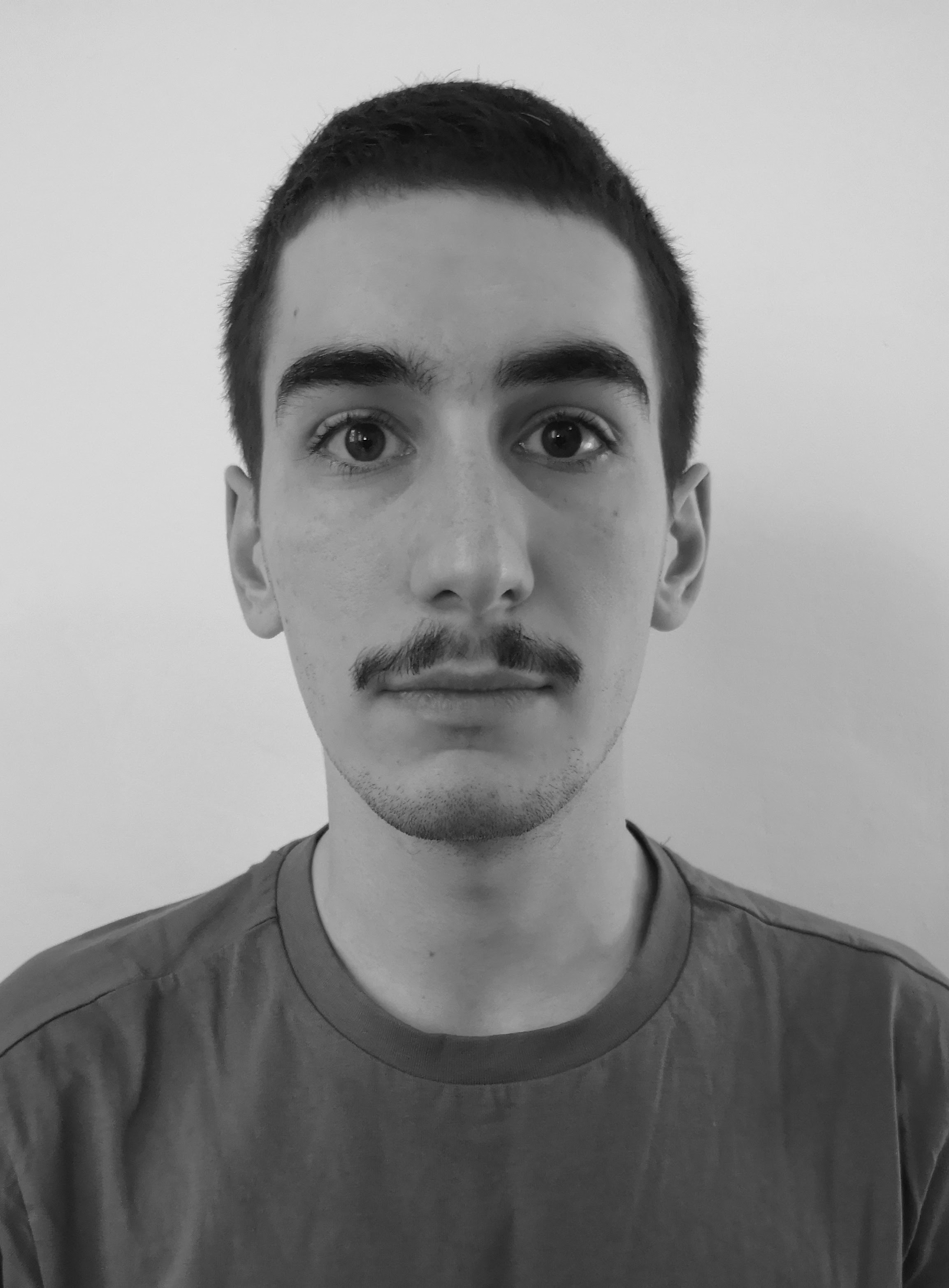}}]{Bence Bak\'o} received his M.Sc. degree in physics from E\"otv\"os Lor\'and University, Budapest, Hungary in 2023, where he is currently enrolled as a PhD student in quantum computing. His research interests include near-term quantum computing and its applications in machine learning.
\end{IEEEbiography}

\begin{IEEEbiography}[{\includegraphics[width=1in,height=1.25in,clip,keepaspectratio]{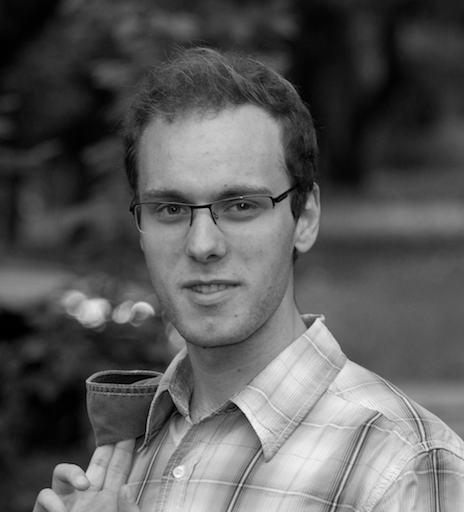}}]{D\'aniel\,\,\,\,\,T.\,\,R.\,\,\,\,\,Nagy} received his M.Sc. degree in physics from E\"otv\"os Lor\'and University, Budapest, Hungary in 2021, where he is currently enrolled as a PhD student specializing on quantum computation and machine learning. His primary focus is quantum machine learning, quantum reinforcement learning and quantum generative modeling.
\end{IEEEbiography}

\begin{IEEEbiography}[{\includegraphics[width=1in,height=1.25in,clip,keepaspectratio]{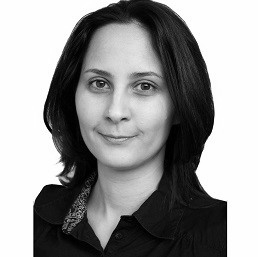}}]
{Zs\'ofia Kallus, Ph.D.} works as a master researcher at Ericsson Research, Traffic Analysis and Network Performance Laboratory in Hungary. She holds a Ph.D. degree in Physics from the Doctoral School of Physics of E\"otv\"os Lor\'and University, Budapest. She studied open quantum systems and the structure and dynamics of large-scale real-world networks. She joined ER in 2014. Currently, she is focused on emerging quantum technologies, quantum computing applications and Trustworthy AI solutions for optimization and automation of high-performance telecommunications systems. She is also interested in sustainability and urban sciences.
\end{IEEEbiography}

\begin{IEEEbiography}[{\includegraphics[width=1in,height=1.25in,clip,keepaspectratio]{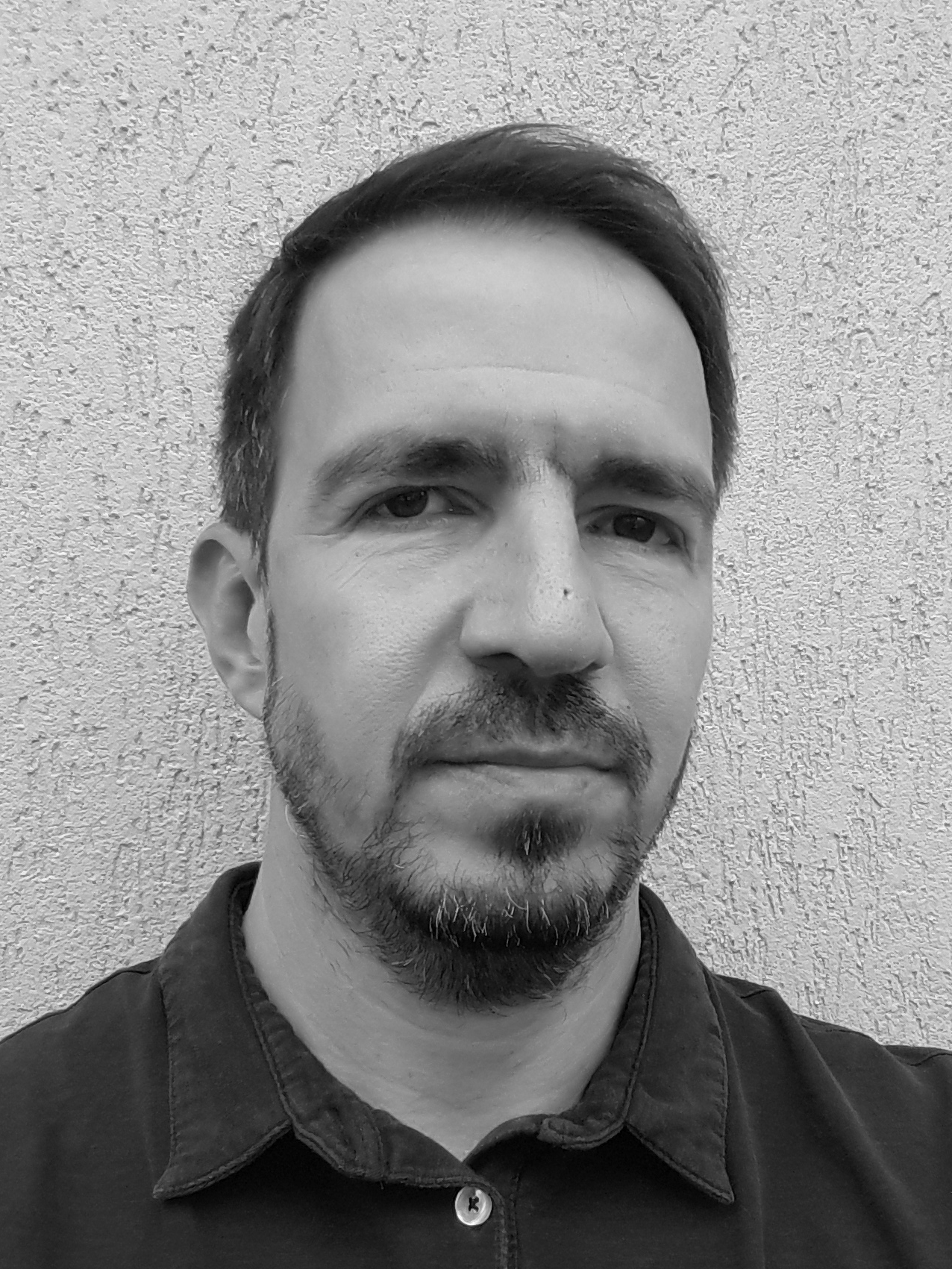}}]{P\'eter Vaderna} works as a master researcher at Ericsson Research, Traffic Analysis and Network Performance Laboratory in Budapest. He received his Ph.D. in physics from the Department of Physics of Complex Systems at E\"otv\"os Lor\'and University, Budapest in 2008, in the area of traffic modeling in communication networks. Currently, he focuses on AI in network analytics, network management and network automation. His research interests also include emerging technologies that can potentially be involved in various business areas of telecommunication such as quantum AI, camera based positional tracking and AR/VR.
\end{IEEEbiography}

\begin{IEEEbiography}[{\includegraphics[width=1in,height=1.25in,clip,keepaspectratio]{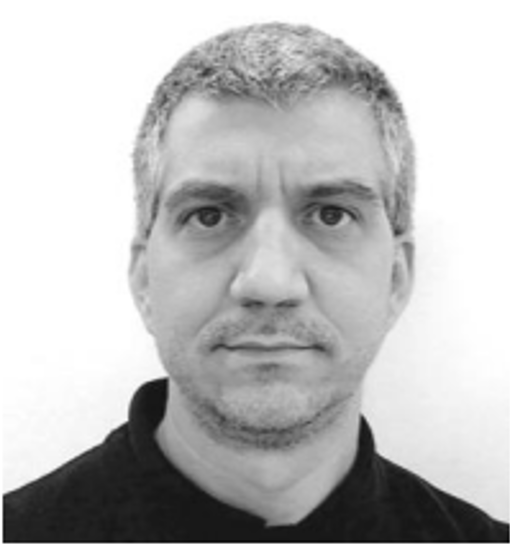}}]{P\'eter H\'aga} 
is a Master Researcher at Ericsson Research. He received his MSc and PhD in physics from the Eötvös Loránd University, Budapest, Hungary. In 2011 he joined Ericsson Research, where he is mostly involved in data analytics and AI development works. He co-authored 30 scientific publications and he is a co- inventor in 20 granted patent applications.
\end{IEEEbiography}

\begin{IEEEbiography}[{\includegraphics[width=1in,height=1.25in,clip,keepaspectratio]{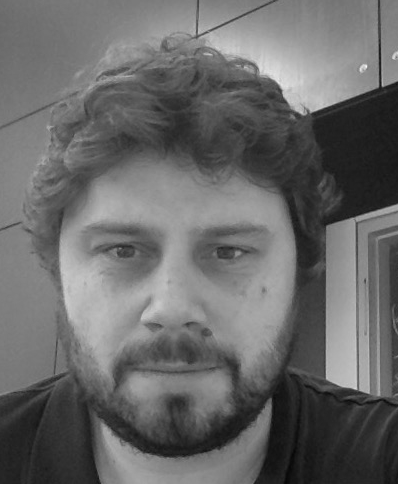}}]{Zolt\'an Zimbor\'as} is the head of the Quantum Computing and Information Group at the Wigner Research Centre for Physics (Budapest). He obtained his Ph.D. at the E\"otv\"os University in Budapest and worked later at ISI Torino, University College London, and Freie Universit\"at Berlin before returning to Hungary. He is a member of several editorial boards and is also a board member of QWorld, a non-profit organization that aims to develop  open-source software  for quantum programming at all levels. His research interests include Quantum Computing, Quantum Control and Quantum Statistical Physics.
\end{IEEEbiography}

\EOD

\end{document}